# Secret Hidden in Navier-Stokes Equations: Singularity and Criterion of Turbulent Transition

Hua-Shu Dou
Faculty of Mechanical Engineering and Automation,
Zhejiang Sci-Tech University,
Hangzhou, Zhejiang 310018, China
Email: huashudou@yahoo.com

**Abstract**

A new formulation of the Navier-Stokes equation, in terms of the gradient of the total mechanical energy, is derived for the time-averaged flows, and the singular point possibly existing in the Navier-Stokes equation is exactly found. Transition of a laminar flow to turbulence must be implemented via this singular point. For pressure driven flows, this singular point corresponds to the inflection point on the velocity profile. It is found that the stability of a flow depends on the direction of the gradient of the total mechanical energy for incompressible pressure-driven flow. When this direction is nearer the normal direction of the streamline, the flow is more unstable. It is further demonstrated that the existence of the singularity in the time-averaged Navier-Stokes equation is the necessary and sufficient condition for the turbulent transition. In turbulent transition, it is observed that the role of disturbance is to promote the flow approaching to produce this singular point. These results are the most important part of the energy gradient theory.

## 1. Introduction

Turbulence is one of the most difficult problems in fluid dynamics. Since the pioneer work for pipe flow carried out by Reynolds was done in 1883, researchers have made much effort in this field [1]. However, the mechanism of turbulence generation is still not fully understood. As is well known, there is a critical Re of about 2000 for pipe flow [2, 3]. When the Re is lower than this critical Re, the flow keeps laminar no matter how of the disturbance. When the Re is larger than this critical Re, the flow may transit to turbulence, depending on the disturbance input [4-8]. Recent studies showed that the formation a turbulent spot in pipe flow is related to the generation of inflection point on the velocity profile [9]. Nevertheless, how the turbulent transition is relevant to the inflection point is not explained in the community. Furthermore, there is no appropriate theory to support the phenomenon of turbulent transition related to the formation of inflection point. It should be mentioned that there was a linear stability theory for inviscid fluid to relate the instability to the inflection point on the velocity profile in Rayleigh [10]. Rayleigh proved mathematically that the existence of inflection point on velocity profile is a necessary condition for instability of inciscid parallel flows [10]. However, since turbulent flow occurs only in viscous flow which is much different from inviscid flow [2-3], it is questionable for an inviscid stability theory to be used to deal with turbulent transition. The most evident features of turbulence are the drag enhancement and energy dissipation caused by fluid viscosity, which are absent in inviscid flow.

Researches showed that the transition from a laminar flow to turbulence is associated with a deformation of the mean velocity profile from a parabolic to a plug profile [11-13]. During the process of transition, besides the momentum transfer, there is also mass transfer in the transverse direction. Since there is no force exerted in the transverse direction, what causes the mass transfer is still not known. Dou and co-authors proposed that the gradient of the total mechanical energy in transverse direction is the source for mass and momentum transfers during the transition for pressure driven flows [14-17]. On other hand, whether there is any change in the characteristic of the Navier-Stokes equations is not clear accompanying the variation of velocity profile. Further, how the velocity profile changes could leads to turbulent transition needs to be further clarified.

As is well known, there is discontinuity from laminar flow to turbulence in the time-averaged Navier-Stokes equations. In other words, singular point may implicitly exist in the Navier-Stokes equations for a given flow configuration. Transition of a laminar flow to a turbulent flow must be via the singular point. However, how the singularity of Navier-Stokes equations is related to the turbulent transition is not understood in the community.

The time-averaged Navier-Stokes equation of incompressible fluid is as follow for laminar and turbulent flows, respectively [2-3],

$$\rho(\frac{\partial \mathrm{u}}{\partial t} + \mathrm{u} \cdot \nabla \mathrm{u}) = \nabla p + \mu \nabla^2 \mathrm{u} . \tag{1}$$

$$\rho(\frac{\partial \mathrm{u}}{\partial t} + \mathrm{u} \cdot \nabla \mathrm{u}) = \nabla p + \mu \nabla^2 \mathrm{u} + \nabla \cdot \tau_\mathrm{t} . \tag{2}$$

where $\tau_\mathrm{t}$ is the turbulent stress tensor. Thus, there is discontinuity from laminar flow to turbulence in the time-averaged Navier-Stokes equations due to existence of $\tau_\mathrm{t}$ (Fig.1). This discontinuity means that there is singularity in the time-averaged Navier-Stokes equations of the laminar flow. Laminar flow to turbulence transition (LFTT) must be implemented via this singularity. Finding out the singularity in Navier-Stokes equations may be useful to reveal the mechanism of turbulent transition. In this study, a new formula of the time-averaged Navier-Stokes equation is derived. The possible singularity existing in the Navier-Stokes equation is found. Then, the mechanism of turbulent transition is discussed.

## 2. Formulation of time-averaged Navier-Stokes equations

For the pressure driven flow between two parallel walls (Fig.2), the continuity and Navier-Stokes equation of the time-averaged laminar flow for incompressible fluid can be written as follow by neglecting the gravitational energy [2-3],

$$\nabla \cdot \mathrm{u} = 0 . \tag{3a}$$

$$\rho(\frac{\partial u}{\partial t} + u \cdot \nabla u) = \nabla p + \mu \nabla^2 u . \tag{3b}$$

and the boundary condition is
$$u = 0. \quad \text{(on the walls)}. \tag{3c}$$

With the identity,
$$u \cdot \nabla u = \nabla(\frac{1}{2}\rho V^2) + \rho(u \times \nabla \times u) . \tag{4}$$

The Eq.(3b) is re-written as
$$\rho \frac{\partial u}{\partial t} + \nabla(p + \frac{1}{2}\rho V^2) = \mu \nabla^2 u + \rho(u \times \nabla \times u) . \tag{5}$$

Here, $V$ is total velocity and $p + \frac{1}{2}\rho V^2$ is the total mechanical energy of unit volumetric fluid.

For laminar flow, the equation for time-averaged flow is the same as that in steady state flow, $\frac{\partial u}{\partial t} = 0$, the Navier-Stokes equation becomes,

$$\nabla(p + \frac{1}{2}\rho V^2) = \mu \nabla^2 u + \rho(u \times \nabla \times u) . \tag{6}$$

Using Eq.(6), the components of the gradient of the total mechanical energy in the transverse direction and the streamwise direction can be expressed, respectively, as [14-17],

$$\frac{\partial E}{\partial n} = \frac{\partial (p + (1/2)\rho V^2)}{\partial n} = \rho(u \times \omega) \cdot \frac{dn}{|dn|} + (\mu \nabla^2 u) \cdot \frac{dn}{|dn|} = \rho V \omega + (\mu \nabla^2 u)_n \tag{7a}$$

$$\frac{\partial E}{\partial s} = \frac{\partial (p + (1/2)\rho V^2)}{\partial s} = \rho(u \times \omega) \cdot \frac{ds}{|ds|} + (\mu \nabla^2 u) \cdot \frac{ds}{|ds|} = (\mu \nabla^2 u)_s \tag{7b}$$

where $\omega = \nabla \times u$ is the vorticity.

Thus, the magnitude and the direction of the gradient of the total mechanical energy for unit volumetric fluid can be written as, respectively,

$$\left|\nabla\left(p + \frac{1}{2}\rho V^2\right)\right| = \sqrt{[(\rho V \omega) + (\mu \nabla^2 u)_n]^2 + (\mu \nabla^2 u)_s^2} \tag{8a}$$

$$\tan \alpha \equiv \frac{\partial (p + (1/2)\rho V^2)/\partial n}{\partial (p + (1/2)\rho V^2)/\partial s} = \frac{\rho V \omega + (\mu \nabla^2 u)_n}{(\mu \nabla^2 u)_s} \tag{8b}$$

where $\alpha$ is the angle between the direction of the gradient of the total mechanical energy and the direction of the velocity vector.

***These two equations are the new form of the Navier-Stokes equation for laminar flow, in terms of the gradient of the total mechanical energy.***

Similarly, using Eq.(2), the time-averaged Navier-Stokes equation for turbulent flow can be written as follow using the same procedure as above,

$$\left|\nabla\left(p+\frac{1}{2}\rho V^2\right)\right|=\sqrt{[\rho V\omega+(\mu\nabla^2 u)_n+(\nabla\cdot\tau_t)_n-(\partial u/\partial t)_n]^2+[(\mu\nabla^2 u)_s+(\nabla\cdot\tau_t)_s-(\partial u/\partial t)_s]^2}$$

(9a)

$$\tan\alpha\equiv\frac{\partial(p+(1/2)\rho V^2)/\partial n}{\partial(p+(1/2)\rho V^2)/\partial s}=\frac{\rho V\omega+(\mu\nabla^2 u)_n+(\nabla\cdot\tau_t)_n-(\partial u/\partial t)_n}{(\mu\nabla^2 u)_s+(\nabla\cdot\tau_t)_s-(\partial u/\partial t)_s}$$

(9b)

Letting $E=p+\frac{1}{2}\rho V^2$ stand for the total mechanical energy, Eqs.(8a) and (8b) read

$$|\nabla E|=\sqrt{[(\rho V\omega)+(\mu\nabla^2 u)_n]^2+(\mu\nabla^2 u)_s^2} \qquad (10a)$$

$$\tan\alpha\equiv\frac{\partial E/\partial n}{\partial E/\partial s}=\frac{\rho V\omega+(\mu\nabla^2 u)_n}{(\mu\nabla^2 u)_s} \qquad (10b)$$

Defining a new variable as $K\equiv\tan\alpha\equiv\frac{\partial E/\partial n}{\partial E/\partial s}$, the Eq.(10b) now is,

$$K=\frac{\rho V\omega+(\mu\nabla^2 u)_n}{(\mu\nabla^2 u)_s} \qquad (11)$$

In above equation, $K$ is a function of coordinates and represents the direction of the gradient of the total mechanical energy (Fig.3), which is dimensionless. It can also be considered as a *local Reynolds number* [14-17]. It is seen that if $(\mu\nabla^2 u)_s=0$, the Eq.(11) is singular and the value of K tends to infinite.

Since *K* in Eq.(11) represents the direction of the gradient of the total mechanical energy for pressure driven flows, which controls the stability of a flow under certain disturbance, this is the origin of the name "*Energy gradient theory*" [14-17].

**Special cases:**

(1) For parallel flows, $(u)_n\equiv 0$, $(\mu\nabla^2 u)_n=0$, and $V=u$, then,

$$K=\frac{\rho u\omega}{(\mu\nabla^2 u)_s} \qquad (12)$$

If $(\mu\nabla^2 u)_s=0$, then $K=\infty$, the equations (11) or (12) is singular. This means that an inflection point on the velocity profile appears.

Generally, the fluid viscosity plays a role of stability in parallel flows, from Eq.(12). At an inflection point of parallel flows, role of viscosity vanishes. Thus, the rate of amplification to a disturbance at the inflection point is infinite, and the laminar flow is able to involve into turbulent flow under this condition at a sufficient high Reynolds number.

(2) For inviscid flows, $|\nabla E| = |\rho V \omega|$, $K = \tan \alpha = \infty$ and $\alpha = 90°$. The direction of the gradient of the mechanical energy is always along the normal direction of the streamline.

(3) When $|\nabla E| = 0$, there are two cases: (a) $\rho V \omega = -(\mu \nabla^2 u)_n$ and $(\mu \nabla^2 u)_s = 0$. This means that $\rho V \omega + (\mu \nabla^2 u)_n = 0$ at the inflection point of the velocity profile, which seldom happens. (b) $\mu = 0$ and $\omega = 0$ ($V \neq 0$). This is the inviscid and irrotational flow, i.e., potential flow.

It is found from Eq.(10) that the larger the angle $\alpha$ in Fig.3, the nearer to the singularity the flow. Therefore, larger angle $\alpha$ would lead to the flow more unstable.

## 3. Discussions

### (1) Necessary and sufficient condition for LFTT

**Proof of sufficient condition:** It is observed from Eq.(10) that no matter of the parallel flow, or the non-parallel flow, as long as $(\mu \nabla^2 u)_s = 0$ on the velocity profile, namely, the inflection point existing, then the Navier Stokes equation is singular. Since this type of singular point can not be eliminated in mathematics, the singularity will cause discontinuity of flow parameter values. Because of the discontinuity of flow parameters, the flow after discontinuity will be in another flow state which must be different from laminar flow. This flow state must be turbulent flow. Therefore, it is obtained that *the existence of inflection point on velocity profile is a sufficient condition of turbulent transition for pressure driven flows*. It is proved theoretically, for the first time, that for laminar viscous flow, inflection point existing is a sufficient condition of turbulent transition, regardless of the parallel flow, or non-parallel flow. This conclusion is under the boundary condition of Eq.(3c), which ensures that there is no work input or output to change the distribution of the total mechanical energy. Therefore, it is restricted to pressure driven flows.

**Proof of necessary condition:** If turbulent transition occurs, the laminar flow before transition and the turbulent flow after transition are two different types of flows. They are controlled by different governing equations for the time-averaged flow, Eqs.(1) and (2), respectively. Thus, there is a discontinuity in the governing equation due to the turbulent stresses and hence the Navier-Stokes equation for laminar flow contains at least a singular point at transition. According to Eq. (10), if the Navier-Stokes equation shows a singularity, no matter of the parallel flow, or the non-parallel flow, there must be at least one point on the velocity profile at which $(\mu \nabla^2 u)_s = 0$, namely inflection point existing on the velocity profile. Therefore, for laminar flow (of course, viscous flow), no matter of the parallel flow, or non-parallel flow, *the existence of inflection point on velocity profile is a necessary condition of turbulent transition for pressure driven flows.* As discussed before the restriction of boundary condition Eq.(3c) is applied. The experimental studies for plane Poiseuille flow and pipe Poiseuille flow have confirmed that velocity profile shows an inflection point on the velocity profile during the transition from laminar flow to turbulence [18-19]. More discussions can be found in [14-17].

**(2) Experimental confirmation of turbulent transition implementation via singularity**

A lot of studies in the past concerned the coherent structure in the outer layer of a turbulent boundary layer. Most people concentrated on hairpin-like vortices as the dominant feature in these flows. More recently, it was found that hairpin vortices are the core which develops into turbulent packets. At high Reynolds number flow, these packets form streamwise train of hairpins which contribute significantly to the Reynolds shear stress and sustain the turbulence [20-29]. As is well known, the formation and development of hairpin vortex accompanies the velocity inflection. These phenomena have been described in the published literature [20-29]. So far no any instance is found that turbulent transition occurrence with the absence of velocity inflection.

Recently, experiment for a pipe flow showed that transition to turbulence is strongly related to the velocity inflection [9]. Hof et al's experiments indicated that the inflection point on the velocity distribution is the core of the turbulent spot [9]. Hof et al argued that the transport of streamwise vorticity given by the cross-sectional average of the product of the magnitude of the axial vorticity and the relative motion to the mean velocity, $\langle |\omega_z|(u_z - U) \rangle$, dominates in sustaining the inflection points, which in turn cause the instability that regenerates vorticity. In their simulation, it was found that the strongest inflection point coincides with the location of the vorticity production, and the turbulent kinetic energy reaches maximum after the inflection point, indicating that turbulence is indeed sustained by inflection point instability.

However, the mechanism behind this phenomenon requires further theoretical interpretation. How the inflection point trigs turbulent transition is not fully understood. There is no theory to explain this phenomenon, and there is no mathematical formulation to express the physics of inflection point which is related to turbulent transition. The present work exactly describes theoretically the physics of turbulent transition resulted from the velocity inflection. Velocity inflection firstly leads to the singularity of Navier-Stokes equations. This singularity causes the discontinuity of flow parameters and finally results in turbulent transition, causing the laminar flow to become another type of flow, i.e., turbulence. During the process of velocity inflection formation, hairpin vortices are accompanied to appear. Thus, hairpin vortices are necessary for the generation of turbulence. Since the velocity inflection and the hairpin vortices are dominated by the time-averaged flow, thus, the time-averaged flow of vortical field controls the variation of flow structure in turbulence, which is called the coherent structure of turbulence, as found by experiments and simulations [30-33].

From above discussions, it is understood that turbulent transition is originated from the variation of the distribution of $K$ in time-averaged flow under the role of disturbance and is formed by the singularity of the Navier-Stokes equation. At the beginning of transition, a turbulent spot corresponds to one (or one group) singular point. In the developed turbulence, there are numerous turbulent spots, which correspond to numerous singular points. Therefore, turbulence actually is composed of numerous singular points.

**(3) Explanation to the scaling laws of disturbance with the Re for LFTT**

Hof et al [5] did experiment for the influence of disturbance on LFTT in pipe flow. It is found that the amplitude of the disturbance scales with the Re as an exponent of -1 for normally injected disturbance (Fig.4). This experimental result is still not fully understood so far. In this study, the experiment phenomenon found can be explained using the time-averaged Navier-Stokes equations.

The area below the scaling curve in Fig.4 is the laminar flow and is controlled by Eq.(1). The area above the scaling curve is the turbulent flow and is controlled by Eq.(2). There is discontinuity from Eq.(1 ) to Eq.(2) since the existence of turbulent stress tensor. For a given Re, with the increase of the velocity disturbance, the flow transits to turbulence when it is across the curve from the bottom to the top. Thus, it can be explained that the experimental curve in Fig.4 corresponds to the formation of inflection point on the velocity profile, i.e., singularity of the time-averaged Navier-Stokes equation.

The scaling curve in Fig.4 symbolizes the jump of flow parameters and indicates the discontinuity of the time-averaged Navier-Stokes equation. Thus, the scaling curve should represent the singularity of Navier-Stokes equation. Therefore, the role of disturbance in experiments from a laminar flow to transit to turbulence is to promote the time-averaged flow approaching the formation of singularity, and then the implement of turbulent transition. More discussions can be found in [14-17].

**4. Conclusions**

The conclusions can be drawn as follow:
1. There is singularity in the time-averaged Navier-Stokes equations for laminar flow at certain condition. Transition of a laminar flow to a turbulent flow must be via the singular point.
2. It is derived that the singular point existing in the Navier-Stokes equation is the inflection point on the velocity profile for pressure driven flows.
3. The stability of a flow depends on the direction of the gradient of the total mechanical energy for incompressible pressure-driven flow. When this direction is nearer the normal direction of the streamline, the flow is more unstable.
4. The existence of the singularity in the time-averaged Navier-Stokes equations is the necessary and sufficient condition for the turbulent transition.
5. The role of disturbance is to promote the flow approaching to produce this singular point and to create the condition of turbulent transition.

These results are the most important part of the energy gradient theory [14-17].

**Acknowledgements**
The author thanks Prof. B. C. Khoo for the comments. This work is supported by Science Foundation of Zhejiang Sci-Tech University (ZSTU) under Grant No.11130032241201.

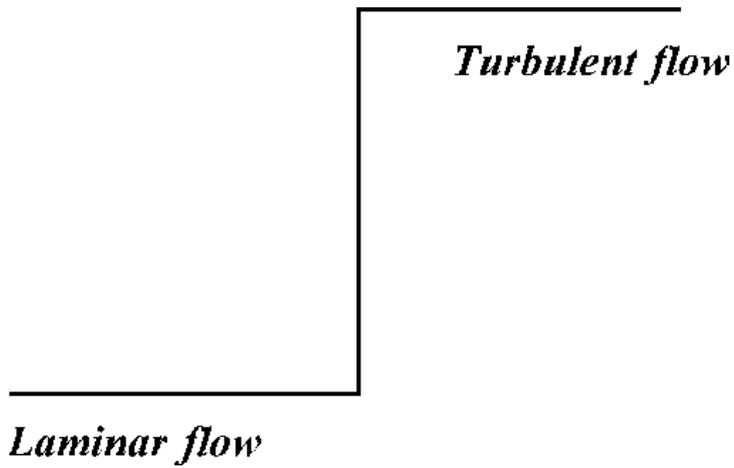

**Figure 1.** Discontinuity existing during transition from laminar flow to turbulent flow. There is a jump for all the flow parameters such as velocity, pressure, total mechanical energy, shear stress as well as drag force, etc.

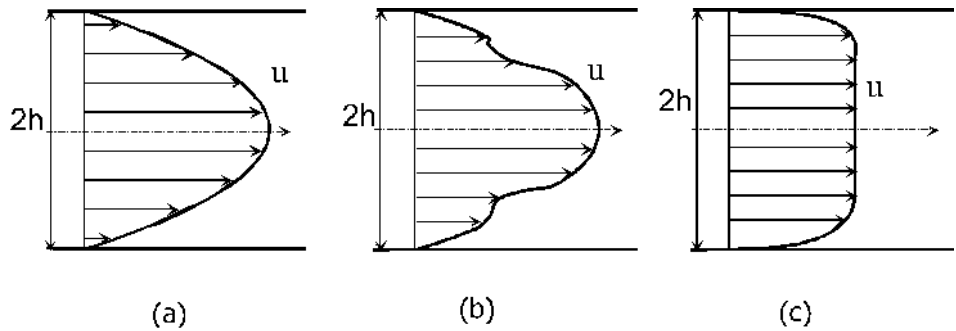

**Figure 2.** Velocity profile during transition from laminar flow to turbulence for plane Poiseuille flow. (a) Laminar flow; (b) Transitional flow; (c) Turbulence.

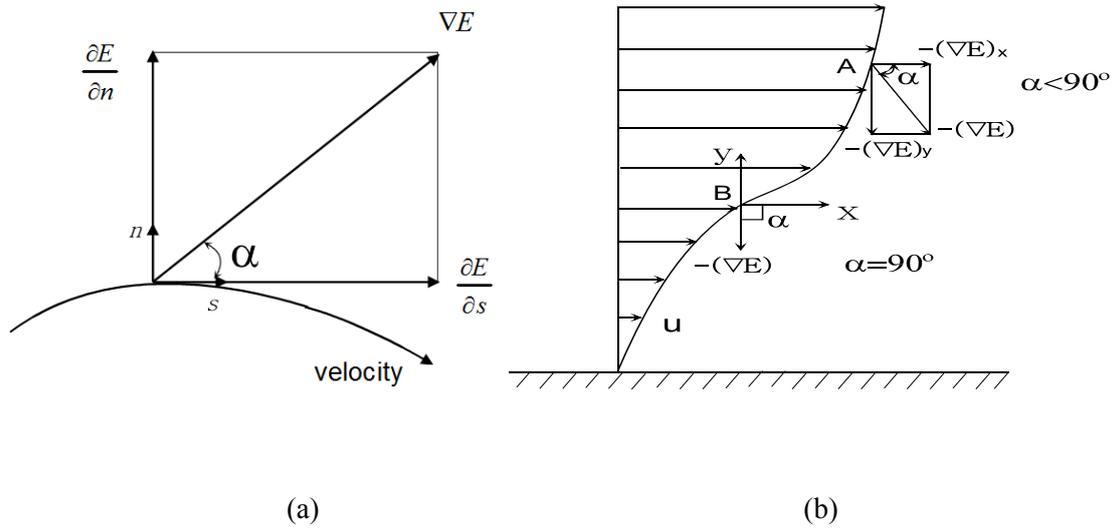

(a)                        (b)

**Figure 3.** (a) Direction of the total mechanical energy gradient; (b) Schematic of the direction of the total mechanical energy gradient and energy angle for flow with an inflection point at which the energy angle equals 90 degree.

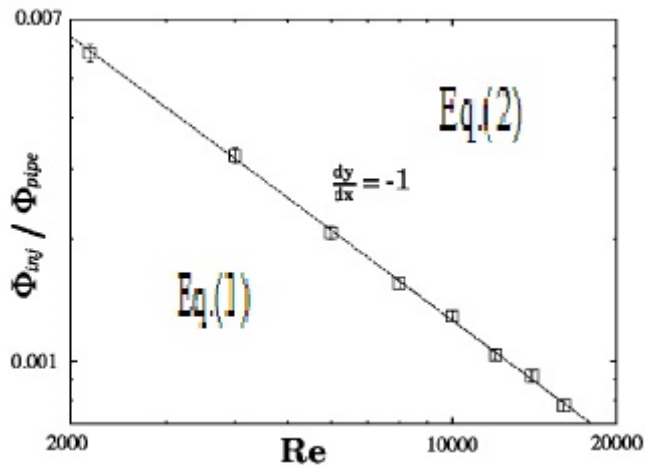

**Figure 4.** Explanation of the scaling law of dimensionless disturbance amplitude with Re found from experiment. Experimental results for pipe flow: the normalized flow rate of disturbance versus the Reynolds number (adapted from Hof, Juel, and Mullin (2003). The range of Re is from 2000 to 18,000. The normalized flow rate of disturbance is equivalent to the normalized amplitude of disturbance for the scaling of Reynolds number, $\Phi_{inj}/\Phi_{pipe} \sim (v'_m/U)_c$ The curve of scaling of disturbance amplitude with the Re represents the singularity of time-averaged Navier-Stokes equations to an extent.